# New dimensions in magnetism and superconductivity via 3D nanoprinting with focused electron and ion beams

**Authors:** Claire Donnelly[1,2,3], Rosa Córdoba[4], Amalio Fernández-Pacheco[5]
[1] Max Planck Institute for Chemical Physics of Solids, 01187, Dresden, Germany
[2] Laboratory for Multidimensional Nanomaterials, Department of Materials, ETH Zurich, 8093 Zurich, Switzerland
[3] PSI Centre for Photon Science, Paul Scherrer Institute, 5232 Villigen PSI, Switzerland
[4] Institute of Molecular Science (ICMol), Universitat de València, València, Spain
[5] Institute of Applied Physics, TU Wien, Wiedner Hauptstr. 8-10/134,1040 Vienna, Austria

## Abstract:
The extension of nanoscale quantum materials to the third dimension brings new possibilities to geometrically tune emergent properties, emulating or indeed going beyond the physics of bulk materials, with exciting implications for high-density technological integration. Realising such three-dimensional architectures requires nanofabrication techniques that not only enable complex 3D geometries, but also deliver functional quantum materials with emergent phenomena.
Recent advances in focused electron- and ion-beam induced deposition (FEBID and FIBID) have made 3D nanopatterning of functional materials in complex geometries increasingly feasible, paving the way for the experimental realisation of new phenomena and device prototyping.
In this review, we discuss the use of FEBID and FIBID 3D nanopatterning of superconducting and magnetic materials. We address the underlying mechanisms, and capabilities of these techniques, focusing on the physical phenomena that become accessible. Considering the resolution of the techniques in the context of the fundamental lengthscales of the materials, we discuss the emergent effects that arise due to three dimensional geometries in magnetic and superconducting nanostructures, and highlight perspectives for future developments of the techniques.

## Introduction:
Patterning quantum materials at the nanoscale offers a route to tailor the emergent properties of a system geometrically at lengthscales well suited for high-density technological integration. While for many years such nanopatterning has been performed in two dimensions using traditional lithographic techniques, there has been a growing interest in the extension to *three-dimensional* nanogeometries[1–6], that promises a change in paradigm for quantum nanomaterials – both in terms of their fundamental properties, and for technological applications. Indeed, the extension to 3D brings new possibilities for geometric control (including chirality[7,8], density[9], and interconnectivity[10].) that can stabilise new textures and lead to new device concepts (discussed in Section 1.2) Realising these prospects experimentally requires the ability to pattern functional materials in complex three-dimensional nanogeometries.
With recent developments in focused electron and ion beam induced deposition (FEBID, FIBID), the 3D nanopatterning of functional materials in complex geometries has become possible[11–13], paving the way for the experimental realisation of phenomena and the prototyping of devices. FEBID and FIBID, which typically make use of commercial scanning electron or focused ion beam microscopes, respectively, involve the injection of a precursor into the chamber, and the local deposition of material with the focused beam, harnessing the precision of the scanned charged beam in combination with computed scanning protocols to achieve complex 3D geometries. These versatile techniques offer a number of advantages: their nanoscale resolution lies at the relevant lengthscales for the magnetic and superconducting materials in question, they are direct write, meaning that the 3D nanostructures can be directly deposited without the need for resists or templates, and on a variety of substrates. Furthermore, the integration with CAD software offers 3D printer-like flexibility in the geometries that can be patterned.
In this review, we discuss the existing capabilities of these techniques and highlight future directions, with a focus on the physical phenomena that can be addressed with the patterning of 3D superconducting and magnetic nanostructures (see **Figure 1**). We begin by discussing the lengthscales, and effects of geometry that emerge in magnetic and superconducting nanostructures. Following this, we describe the different materials that are available, and those we foresee being patternable in the near future. Finally, we conclude the review with a discussion of future developments, in terms of materials, resolution, and scalability. Alongside these main discussions, Box 1 provides an overview of the physical mechanism underlying 3D nanoprinting with FEBID/FIBID (mechanism and growth control), while Box 2 presents an overview of characterization techniques for 3D nanostructures.

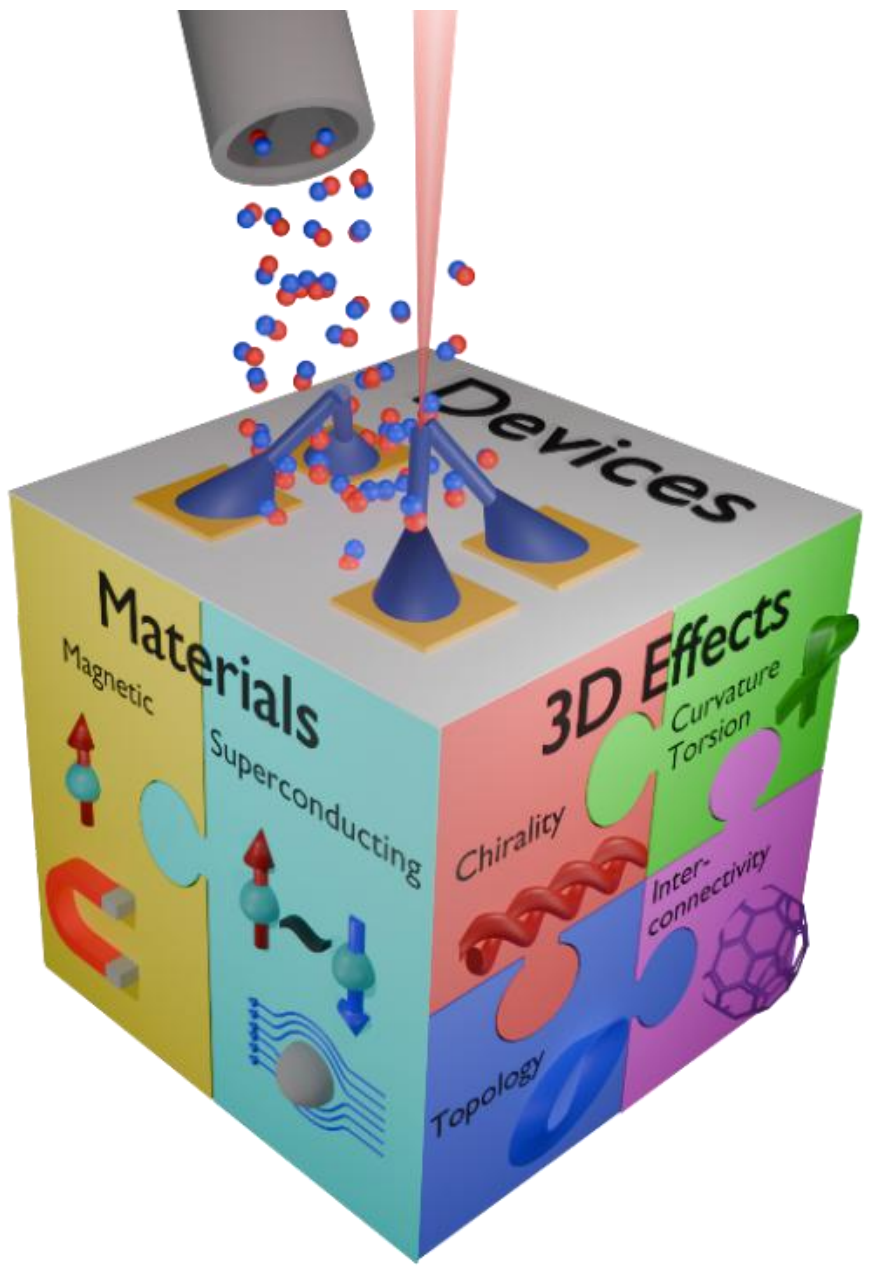

Figure 1. Focused electron and ion beam induced deposition makes possible the creation of 3D magnetic and superconducting nanostructures, incorporating geometrical effects such as chirality, curvature, topology and interconnectivity that can significantly influence the system's properties.

## Main Text:

## 1. Physics of three-dimensional nanogeometries.

When patterned on relevant lengthscales, the nanostructuring of materials can lead to confinement effects that impact the emergent properties of a system. This has been explored for many years in two dimensions: the extension to three dimensions offers new possibilities for both physical phenomena and functionalities[1,3,6].
For example, even when going from a flat nanostrip patterned in two dimensions to a cylindrical nanowire or nanotube that is typically the building block of FEBID/FIBID nanoprinting, the nature of the confinement changes dramatically. In the case of a nanotube, the geometry is topologically distinct from a thin film, which leads to e.g. the formation of different types of magnetic domains. The magnetisation may therefore align axially or azimuthally around the tube, which in turn can result in non-reciprocity in magnon propagation[14]. For superconductors, under the application of a magnetic field the curved nanotube surface hosts not one but two Meissner currents[15]. In the case of a nanowire, the cylindrical geometry can lead to the stabilisation of new magnetic textures[16], the curved cylindrical surface can lead to controlled curvature in superconducting vortices[17], strong confinement effects, as well as rotational symmetry about the long axis of the nanowire impacting anisotropies of the system[17,18].
When one considers the complex three-dimensional nanostructures that can be realised with 3D nanoprinting techniques discussed in this review (see Figure 2), there are a number of opportunities for exploring emergent properties, and fine control thereof. We will first discuss the achievable lengthscales for fabrication with FEBID and FIBID, and how this leads to confinement effects that impact both magnetic and superconducting properties, before going on to discuss the 3D geometric effects themselves.

### 1.1 Confinement effects with sub-100 nm spatial resolution

When considering a fabrication technique, it is critical to consider the achievable minimum feature size in the context of the physically relevant lengthscales of the material or behaviour in question. These fabrication limits are determined by the spatial resolution – i.e. the minimum reproducible feature size – for example the diameter of nanowires or the spacing between structures. In this regard, FEBID and FIBID offer direct-writing of 3D nanoarchitectures with features smaller than 100 nm[19,20], and in favourable conditions with resolutions down to a few nanometres. The achievable resolution is constrained by factors such as the probe size (~1 nm for FEB, 1–7 nm for FIB), the effective beam-matter interaction volume, and surface growth kinetics[21,22]. For instance, nanowire diameters typically range from 30–100 nm, with potential for further miniaturization through post-treatment methods[23].
These nanoscale resolutions can moreover be combined with the direct-writing of complex 3D nanogeometries, including structures with intricate curvature and torsion, such as nanohelices[24] or Möbius strips[12]. The techniques also allow for controlled separation between features of less than 10 nm[25,26], essential for creating high-density 3D nanoscale arrays. Moreover, FEBID and FIBID offer tunability in the composition and microstructure of individual nanostructures by modulating experimental parameters, including electron/ion dose, dwell time, and precursor flux. It must be noted, however, that while FEBID and FIBID offer precise nanoscale patterning of these complex geometries, maintaining fidelity at sub-100 nm scales (in extreme curvatures or highly complex designs) remains challenging, often leading to tapering and local overgrowth, and ensuring uniform properties, across the full 3D volume remains a limitation.

When we consider these lengthscales in the context of materials, we need to consider the physics of magnetic and superconducting systems. FEBID has proven particularly effective for direct-writing 3D magnetic nanostructures[2,6,27]. The resolution of the technique aligns well with relevant length scales of magnetic materials, such as cobalt and iron, including the exchange length (~5–10 nm), dipolar length (~30 nm), and the – slightly larger – domain wall width. The nanowire diameter is a critical parameter influencing magnetic behaviour. Typical diameters between 30–100 nm accommodate magnetic configurations dominated by dipolar interactions, while sub-30 nm structures, achievable through post-processing, approach the exchange length, leading to a higher degree of confinement. Although most wires in this diameter range typically form an axial magnetisation promoted by the shape anisotropy, the degree of confinement determines the type of domain wall that can occur: narrower nanowires host transverse domain walls, while larger diameter wires host Bloch point domain walls that contain a magnetisation singularity[28]. In addition, the separation between nanostructures directly impacts the magnetostatic interaction between structures: indeed, FEBID and FIBID resolutions allow for separations below ~30 nm, enabling fine control of inter-element magnetic interactions, crucial for high-density magnetic devices or logic elements. The FEBID/FIBID resolution is ideally suited to the length scales of these magnetic phenomena, offering a powerful platform for designing and investigating complex 3D magnetic systems.

In the case of 3D superconducting structures, FEBID and FIBID offer excellent opportunities to explore superconducting phenomena at the nanoscale, such as inhomogeneous magnetic fields, coexistence of vortex and vortex-free regions, and topological complexities in Meissner screening currents[17,18,29]. These nanostructures, with dimensions/diameters between the typical superconducting coherence length, $\xi$ (~5–10 nm), and the London penetration depth, $\lambda$ (~450–850 nm), of FEBID/FIBID materials such as tungsten carbide, operate in an intermediate regime, exhibiting distinct field-dependent effects arising from their size and 3D geometry[18,29]. While challenging experimentally, FIBID has been demonstrated for the fabrication of structures with wall thicknesses of ~20 nm and curvature radii as small as ~50 nm[29]. The control on such lengthscales becomes particularly important in hybrid architectures, such as superconductor-ferromagnet (SC/FM), or superconductor-semiconductor (SC/FM), as discussed in Section 1.2.

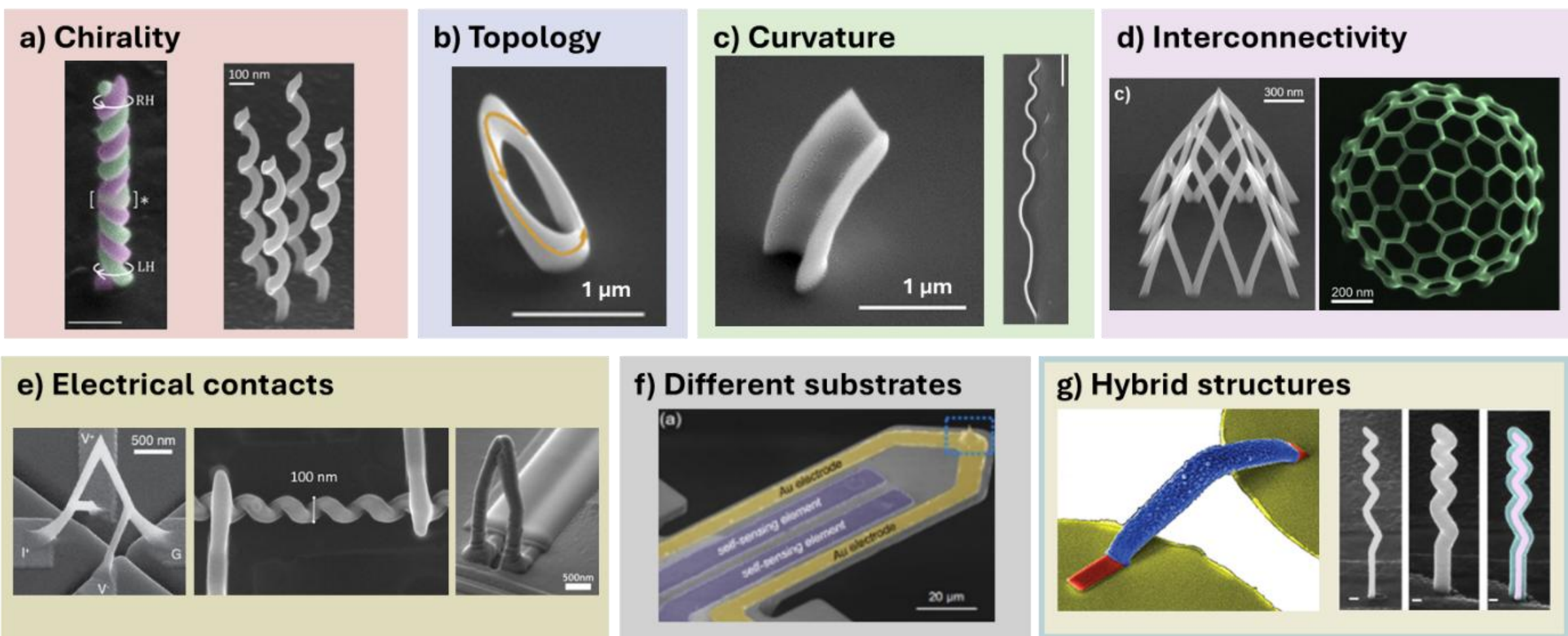

Figure 2. Variety of nanostructure architectures achievable with FEBID/FIBID 3D nanoprinting. With the nanostructures designed with CAD software, it is possible to create a wide variety of geometries, including the definition of a) chiral, b) topologically non-trivial, c) curved and d) interconnected architectures. As direct-write techniques, FEBID/FIBID nanoprinting can be used to directly integrate functional nanostructures into e) electrical contacts, as well as f) patterning directly onto a variety of substrates. Finally, the combination with postprocessing techniques such as chemical vapour deposition and electrodeposition allows for the realisation of hybrid structures. a) reproduced from [18,30]. b,c) reproduced from [12,28]. d) reproduced from [11,142]. e) reproduced from [18,77,93]. f) reproduced from [143]. g) reproduced from [58,59].

### 1.2 Emergent effects arising from three-dimensional geometries

Having discussed the influence of nanoscale confinement that is possible with FEBID and FIBID, we now consider the different geometries and associated phenomena that can be realised with 3D nanoprinting by FEBID/FIBID. In some cases, such as for chirality and curvature, there is already experimental proof of a range of new physical effects, while in other cases, such as non-trivial topologies, there are first experimental demonstrations of the realisation of the necessary geometries for new effects to emerge. In all cases, we will present the state of the art and highlight prospects for the geometric tailoring of phenomena in the future (see **Figures 2-4** and **Box 2**).

***Chirality:***

In condensed matter, chirality is normally considered in the context of broken inversion symmetry. For example, in chiral magnetic materials the absence of inversion symmetry in the unit cell leads to the Dzyalozinskyii-Moriya interactions (DMI), which promotes the formation of helical or topologically non-trivial magnetic textures. In nanostructures, the patterning of chiral geometries such as helices allows one to emulate such chiral behaviours[8], but without the need for materials with intrinsic DMI or symmetry breaking.

In magnetism, FEBID cobalt double helices have been investigated in two regimes. First, overlapping helices, where the two helical strands are connected, and thus interact via the exchange interaction, have been shown to induce a helical spin spiral domain wall, the chirality of which matches the patterned structure[30]. The patterning of so-called artificial chirality on the nanoscale also allows for the investigation of controlled chiral interfaces, at which the chiral region changes handedness. Indeed, it has been observed that topological magnetic configurations such as vortices and Néel defects can be stabilised at these interfaces, as predicted for chiral materials[31], offering a new route to control nanoscale textures. As well as exchange-coupled helices, the precise spatial resolution of FEBID nanoprinting allowed for the realisation of intertwined helices that are spatially separated by distances on the order of 20 nm, leading to the helices being strongly coupled via magnetostatic interactions. Such systems have been shown to host highly coupled domain wall pairs that lead to nanoscale topological textures in the free-space magnetic induction. This concept has recently been extended to lattices of interwoven nanowires, in which arrays of textures in the magnetic induction can be achieved[32]. So far, the vast majority of work has concerned static configurations. However, the fabrication of similar helical geometries with two-photon lithography and atomic layer deposition has led to experimental realisation of a geometric chirality-induced non-reciprocity in magnonics[33]. With first dynamic measurements in 3D FEBID cobalt nanostructures indicating promising properties for such phenomena[34], the exploration of such phenomena in 3D chiral FEBID cobalt nanostructures offers a route to further advance our understanding.

The exploration of chiral geometries for 3D superconducting nanostructures is equally motivating: indeed, as well as chiral geometries leading to new physical phenomena, such helical 3D geometries have been shown to lead to highly efficient sensing of microwave radiation[35]. In terms of 3D nanoprinting techniques, so far, helium-FIBID deposition has been used to fabricate individual and dense arrays of 3D superconducting helices[18]. In these systems, precise control of the helical pitch and radius – and therefore the curvature and torsion – was demonstrated. The chiral geometry was shown through a combination of transport and simulations to lead to localised formation of superconducting vortices, and the formation of so-called phase slips, that are associated with sharp steps in resistance[18]. So far, the vast majority of works have involved the so-called "intermediate regime" of superconductivity, in which the patterned lengthscales are larger than the coherence length, and below the penetration depth. Harnessing such chiral geometries to explore, for example, geometry-induced phase winding, asymmetric supercurrent distributions, modifications of the order-parameter, and symmetry breaking in the superconducting response, requires advances in the growth of materials with longer coherence length.

***Curvature/ torsion:***

Transforming from a flat to a curved space can lead to the introduction of emergent effects across a number of material systems, including both magnetic and superconducting materials. In magnetism, the introduction of curvature and torsion results in new effective energy terms, associated with curvature-induced anisotropy and chirality that arise from the exchange or magnetostatic interactions irrespective of the material. First observations of the effect of curvature on magnetic textures involved 2D systems, in which the propagation and pinning of transverse domain walls was observed [36]. The deposition of thin films on scaffold structures led to first insights into the influence of curvature in curved shells, albeit limited to relatively lateral geometries, in which local curvature control of textures such as magnetic vortices [37] could be achieved. With FEBID/FIBID nanopatterning, one has the possibility to extend such concepts to three dimensions, and for geometrical parameters to be tuned spatially in 3D. In this way, evidence of chirality-dependent depinning in curved 3D scaffold structures consisting of a permalloy thin film deposited on a 3D FEBID platinum ramp has been realised[38]. In addition, in cobalt FEBID nanowires, regions of varying curvature have been exploited to tailor the energy landscapes of three-dimensional magnetic domain walls that contain Bloch point singularities[28]. This enables the definition of pinning points and the realisation of non-reciprocal motion. Similarly, in helical FEBID cobalt structures, the local curvature has been found to determine the magnetisation reversal protocol[39].

The definition of curved superconducting surfaces has been predicted to lead to so-called "geometric forces", in which vortices are attracted towards regions of negative Gaussian curvature[40]. This reflects a curvature-dependent contribution to the vortex line energy on a curved manifold, and is complemented by predictions of superconductivity being induced on the surface of a curved topological insulator. Moreover, in bent nanowires, there exist prospects for the onset of topological superconductivity[41] and spontaneous Josephson effects[42].

Experimentally, the fabrication of three-dimensional superconducting nanostructures with FEBID and FIBID has so far been limited to radii of curvature (or geometric variation lengthscales) on the order of the penetration depth, rather than the coherence length. At these lengthscales, the intrinsic curvature of the surface of the cylindrical nanowires may lead to the curvature of superconducting vortices, as demonstrated in a bridge-like structure[17]. In addition, curved surfaces lead to modulations in the local effective field and promote the coexistence of SC and N states, leading to phase slips[18].

Looking ahead, the precise definition of curved three-dimensional geometries on lengthscales corresponding to both the coherence and penetration depths will offer a route to the manipulation of superconducting phenomena in 3D.

***Topology:***

Introducing topologically non-trivial geometries can impact emergent effects. In the simplest case for soft magnetic materials such as cobalt or iron, going from a wire to a torus, one promotes the change from a single dipole to a vortex state[43] which, combined with intrinsic effects such as chirality, can lead to additional emergent states such as magnetic hopfions[44] or topologically non-trivial helical configurations[45]. Extending to topologically non-trivial geometries such as a Möbius strip or trefoil knot can lead to non-trivial domain wall behaviour[46], as well as the stabilisation of knotted magnetisation patterns. Such structures can also host strongly coupled regions hosting domain walls that are highly interacting but remain unable to annihilate.

For superconductors, the exploration of non-trivial topological geometries is equally compelling. Even the simple extension of a straight wire into a torus enables the formation of so-called "persistent currents". These are dissipationless equilibrium currents associated with fluxoid quantisation in a closed superconducting loop. In a Möbius strip, a non-trivial phase diagram has been predicted, including the existence of nodal states, *i.e.* regions where the order-parameter amplitude vanishes due to Möbius boundary conditions and phase frustration [47].

To date, however, there have been relatively few experimental studies of three-dimensional topological geometries largely due to the challenges associated with their fabrication. Nevertheless, the successful fabrication of a nanoscale magnetic Möbius strip with FEBID[12] demonstrates the ability of the techniques to realise such structures.

***Interconnectivity:***

One of the most important advantages of moving to three dimensions is the prospect of achieving higher interconnectivity, which is crucial for the development of non-traditional computing architectures such as neuromorphic computing. With FEBID/FIBID 3D nanoprinting, the realisation of such interconnected structures has been demonstrated across a range of architectures[17,48–52].

In magnetism, such interconnected systems have been explored in the context of magnetic frustration: the ability to pattern in 3D allows for the mimicking of the true geometry of spin ice materials[53]. First experimental studies have harnessed FEBID nanopatterning to create the building blocks of artificial spin ice, including tripod, tetrapod and nanocube structures[48,50]. As well as determining the overall magnetic configuration of such building blocks, first insights into the three-dimensional magnetic configuration of tetrapod structures - the building blocks of interconnected arrays – have revealed the presence of topological defects at the vertices, that are associated with a non-trivial magnetic vorticity[50]. In such a nanoarchitecture, the combination of nanowires of different orientations also allows the legs to be individually switched with magnetic fields – making possible the field-driven reconfiguration of the magnetic state[51].

In superconductivity, the prospect of realising interconnected architectures is particularly motivating when it comes to phenomena that emerge from the coexistence of different states. Indeed, a recent demonstration of a three-dimensional superconducting nanoarchitecture formed of interconnected nanowires fabricated with FEBID W-C[17] was shown to host both superconducting and normal states, determined by local variations of the orientation of the structure to the applied field. In particular, the normal state could be induced geometrically in both nanowire segments and localised vertices, thus establishing a possible route towards geometrically reconfigurable Josephson junctions. Since planar Josephson junctions and superconducting quantum interference devices have already been demonstrated using FEBID[54] and FIBID[55], extending these concepts to 3D opens a route towards three-dimensional superconducting sensing devices.

***Hybrid / heterostructures:***

Beyond the fabrication of single component nanostructures, the prospect of hybrid systems comprising multiple materials offers the opportunity to go beyond the nanowire and volume geometries that FEBID/FIBID are naturally suited to producing. Such heterostructures have been explored in a number of ways.

First, FEBID and FIBID have been used to create scaffold structures, which are then functionalised with thin film deposition techniques. In this way, shell-like coatings of soft ferromagnets such as permalloy can be created with anisotropic thin film deposition techniques such as evaporation, which have been used to explore the propagation of domain walls in ramp-like[38] and curved conduits[56]. As demonstrated for a simple bridge-like structure, this scaffold approach can be generalised to allow the integration of complex thin film heterostructures into electrical circuits[57].

In order to create nanotube-like structures, the deposition of conformal coatings is key. This has recently been achieved via a local chemical vapour deposition (CVD) process induced by Joule heating, in which a current is passed through a contacted FEBID nanostructure in the presence of a precursor[58], enabling thermally-driven conformal growth. In this way, both magnetic ($Co_3Fe$), and superconducting (NbC) nanotube structures have been realised. In addition, conducting FEBID-deposited materials have been exploited to realise conformal electrodeposition of magnetic nickel shells on 3D FEBID scaffolds[59], leading to the realisation of three-dimensional magnetic nanotubes. Finally, it has also been possible to deposit core-shell heterostructures, including metal/insulator/semiconductor structures[60] and core-shell ferromagnetic nanowires[61] using purely FIBID/FEBID.

While most studies of 3D-grown heterostructures have so far focused on the functionality of a single component such as ferromagnetic nanotubes, the possibility of fabricating complex hetero-nanostructures opens new directions in superconducting spintronics. In particular, interfaces between superconductors and ferromagnets are known to enable the generation of spin–triplet pairs via magnetic inhomogeneities[62–64]. Remarkably, it has been predicted that such triplet components could also be induced via geometric curvature[65,66], acting as an effective source of spin mixing by locally modifying the pairing symmetry.

To realise such phenomena, interface quality and spatial separation become critical. For instance, in SC/FM or SC/SM hybrid systems, coupling across distances on the scale of the coherence length $\xi$ can lead to Josephson effects, spin–triplet supercurrents[67], or superconducting spin valves[68,69]. Moreover, at nanometric separations comparable to the vortex core size (~10 nm), vortex–vortex interactions may dominate[29]. Encouragingly, high-quality interfaces have already been demonstrated via FEBID in magnetic tunnel junctions[70], highlighting that such complex interfacial phenomena may become accessible in 3D SC/FM architectures.

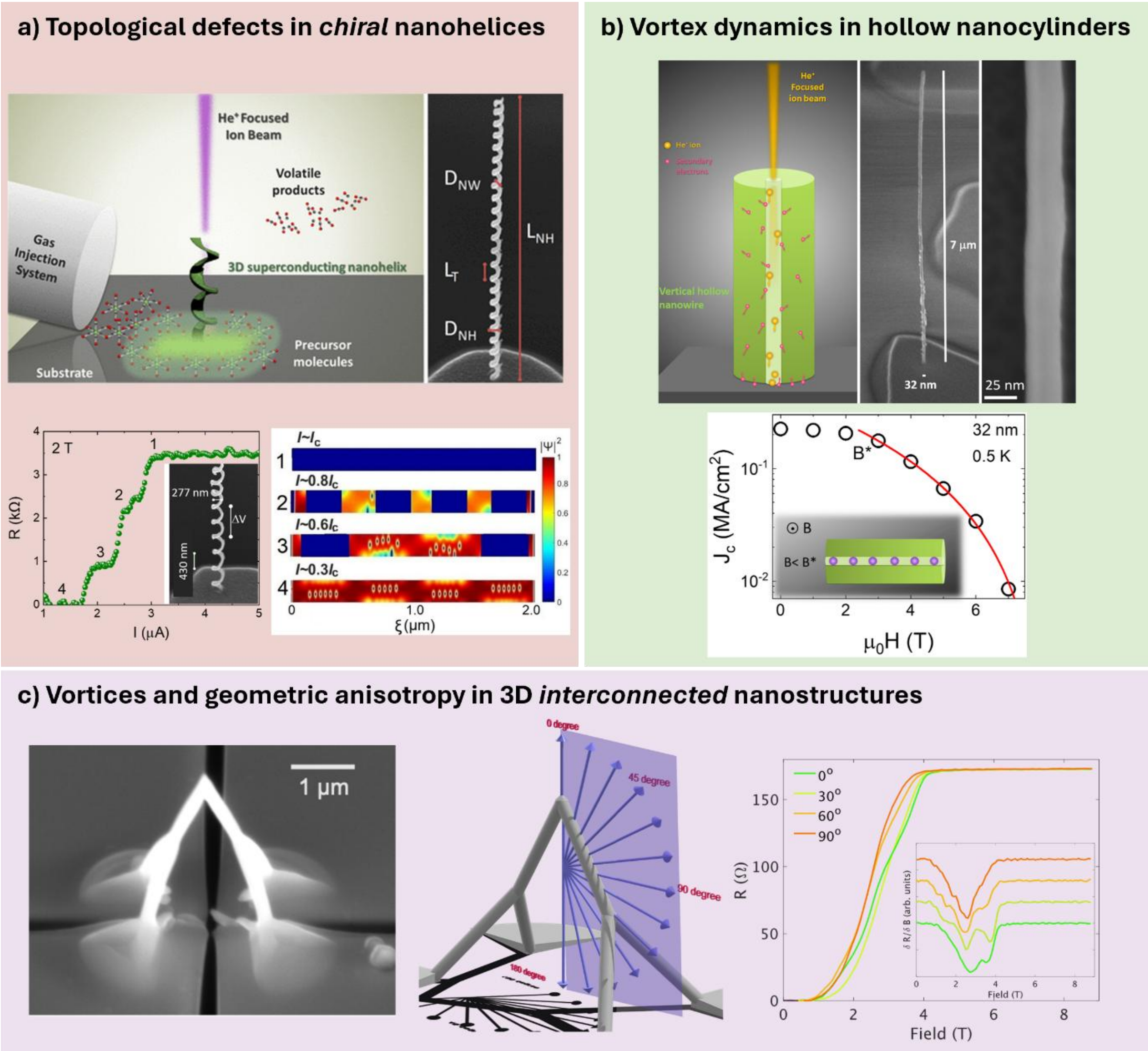

Figure 3. Physics of 3D superconducting nanostructures. a) In superconducting nanohelices, the presence of phase slips that occur due to the inhomogeneous component of the applied field parallel to the nanowire surface is detected via transport measurements, and confirmed with time dependent Ginzburg Landau simulations. b) In superconducting hollow nanocylinders, the presence of a ~6 nm-diameter hole at the centre of the standing cylinder, arising from the balance between milling and deposition processes, leads to a characteristic crossover in $J_c(B)$ revealed by transport measurements. Below B*, the behaviour of $J_c$ is governed by the strong confinement potential of a single row of vortices pinned at the low-energy core of the hollow nanocylinder. c) a 3D interconnected superconducting nanostructure fabricated with FEBID exhibited geometric anisotropy of the critical field, due to inhomogeneous components of the applied field with respect to the nanowire surface. a) reproduced from [18]. b) reproduced from [29]. c) reproduced from[17].

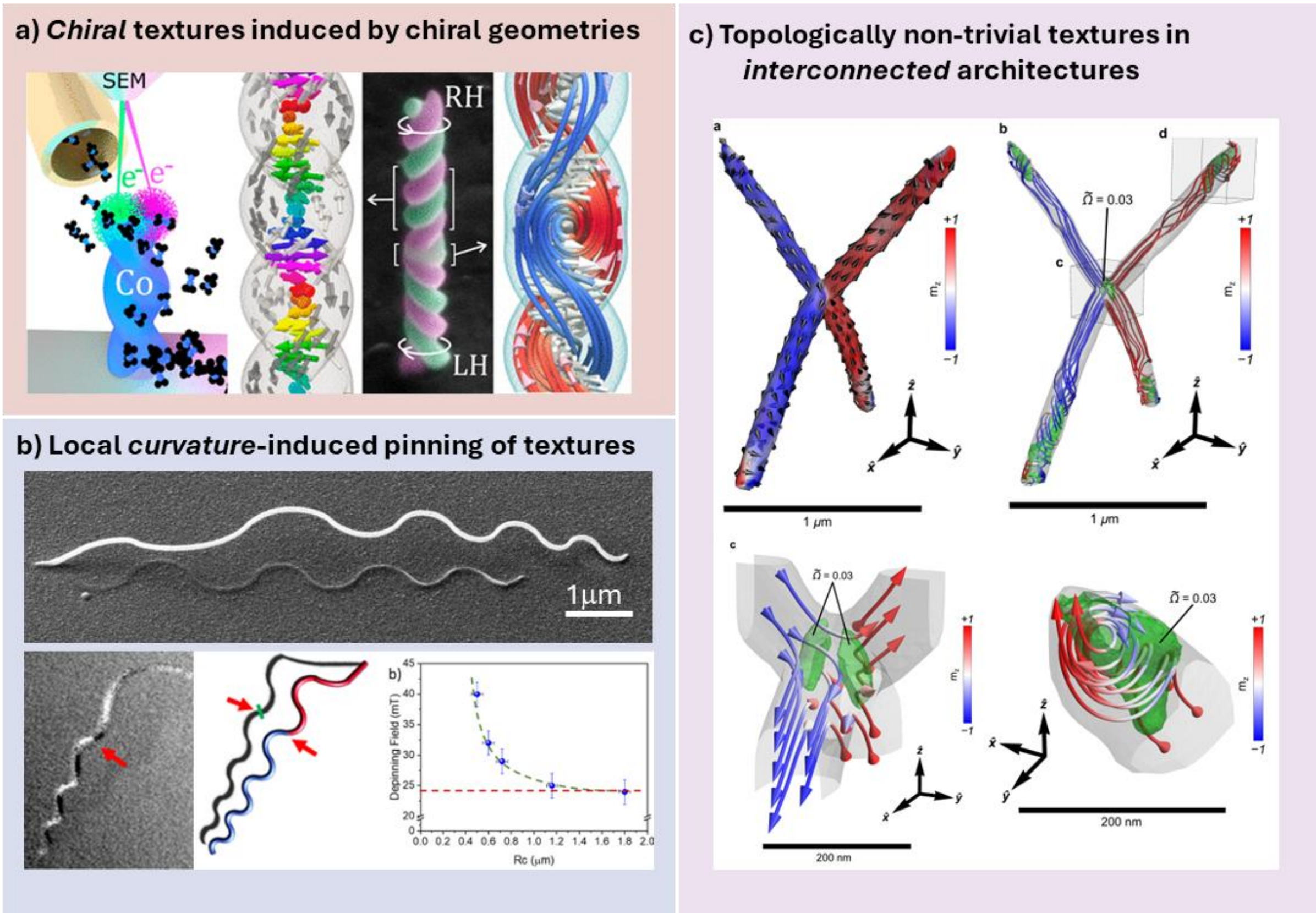

Figure 4. Physics of 3D magnetic nanostructures. a) the patterning of a magnetic double helix with FEBID results in a chiral domain wall forming at the intersect between the two helices, with the chirality determined by the patterned geometric chirality. b) the introduction of curvature into a cylindrical magnetic nanowire modulates the energy landscape of Bloch point domain walls, with regions of high curvature acting as energy barriers for Bloch-point domain wall motion within the nanostructure. c) The definition of interconnected nanowires results in the presence of topologically non-trivial magnetic textures at the vertices. a) reproduced from [30]. b) reproduced from [28]. c) reproduced from [50].

## 2. Direct-write functional materials.

As well as offering a wide variety of geometries, FEBID and FIBID can also be applied to a variety of materials that open up different scientific directions. In this sub-section we provide an overview of the different functional materials that can be grown in 3D, with a particular focus on those that exhibit magnetic or superconducting properties. These materials include cobalt, iron, binary and ternary compounds, tungsten, niobium, molybdenum, and lead, with a complete list given in **Table 1**.

*Magnetic materials*

The fabrication of three-dimensional magnetic nanostructures using FEBID and FIBID has witnessed remarkable progress in recent years[2,71]. Key materials include cobalt (Co), iron (Fe), cobalt-iron (Co-Fe), and iron-platinum (Fe-Pt). Initial studies focused on the generation of cobalt nanostructures with simple geometries such as high-purity nanopillars[72,73], which were used as high aspect ratio magnetic force microscopy (MFM) tips[74]. Insight into the magnetic configuration and switching behaviour was obtained with electron holographic tomography[75], and field-dependent magneto-optical Kerr effect (MOKE) experiments[76]. More recently, advances in 3D FEBID have enabled the creation of more complex geometries such as Möbius strips, double helices, scaffold bridges, curved nanowires, and even a nanoscale replica of a human hand[12,28,30,56,57,77–79]. These intricate 3D geometries facilitate precise control over magnetic chirality, domain wall configurations, and magnetostatic interactions, unlocking new opportunities in spintronics and nanomagnetic logic applications.

In addition to cobalt, there have been a number of studies on iron nanostructures fabricated via FEBID, which range from 2D dots, rods and square frames to 3D nanopillars, all of which exhibit high quality magnetic properties[80–82]. By varying the dimensions and aspect ratios of these structures, it has been possible to define domain-wall pinning sites and to transition between single-domain and flux-closure multi-domain states.

The extension to magnetic binary alloys has led to a significant expansion of FEBID capabilities. Co-Fe wireframe geometries, such as nano-trees, cubes, tetrapods, semi-pyramids, diamonds, and scaffold microbridges[50,58,83], as well as Fe-Pt nanorods[84], have been shown to exhibit robust ferromagnetic properties. Their tunable characteristics and complex geometries make them promising candidates for high-performance magnetic devices.

While the materials utilized so far in 3D remain somewhat limited, there have been a much wider range of materials demonstrated in 2D deposition. These advances provide clear opportunities for extending the accessible material palette to more complex 3D architectures. In particular, Co-Pt-C[85] and Co-Si-C[86] alloys are particularly relevant because they may exhibit intrinsic (bulk) chirality, arising from their non-centrosymmetric atomic scale structure. Patterning these materials in 3D could offer the possibility to combine geometrical chirality with the intrinsic DMI, as has been recently demonstrated through the 3D nanopatterning of bulk chiral magnets[45,87]. A number of additional materials containing magnetic elements, including Co-F[88], Ni-O[89], Fe-Si[90], and $Co_2FeSi$[91], have also been successfully deposited in 2D with FEBID, though their magnetic properties remain unconfirmed. Finally, compounds such as Fe-Si and $Co_2FeSi$ Heusler alloys have also been produced through sequential FEBID deposition and electron beam-induced atomic intermixing, opening the door to the combination of complex geometries with non-trivial material symmetries.

To summarise, ten magnetic materials can now be fabricated using FEBID and FIBID, with the 3D nanopatterning of Co, Fe, Co-Fe, and Fe-Pt having been demonstrated at sub-100 nm lengthscales. With a number of exciting prospects to develop the 3D nanodeposition of more complex materials, FEBID and FIBID will continue to play a central role in the growing field of 3D nanomagnetism, to make possible complex magnetic structures with novel emergent properties.

*Superconducting materials*

In the case of superconducting systems, the most widely used material for 3D nanopatterning via FEBID/FIBID is tungsten-carbon (W-C). Indeed, with this precursor, 3D nanostructures have been successfully fabricated using both $Ga^+$ and $He^+$ FIBID, as well as FEBID. To begin with, $Ga^+$ FIBID has been employed to produce W-C wires[92–94], achieving critical temperatures ($T_c$) and upper critical magnetic fields ($B_{c2}(0)$) that are comparable to those of planar nanostructures, ≈5 K and ≈9 T, respectively. In contrast, $He^+$ FIBID has demonstrated the ability to create more complex geometries, such as hollow nanocylinders[29,95] and nanohelices[18], which exhibit enhanced superconducting properties ($T_c \approx$ 6.2−7.1 K, $B_{c2}(0) \approx$ 12−15 T). For instance, the fabrication of hollow nanocylinders has been demonstrated, with diameters as small as 32 nm, aspect ratios close to 200, and wall thicknesses as low as 13 nm. Similarly, nanohelices with diameters of ~100 nm combine high critical fields with distinctive phase slips, which are commonly attributed to topologically non-trivial screening currents and strong confinement effects arising from the 3D geometry. Recently, first demonstrations of the deposition of 3D superconducting WC nanostructures with FEBID offer precise control for fabricating more complex geometries using 3D nano-printing algorithms[17]. This allowed for the realization of a 3D bridge-like structure, in which superconducting vortex motion, as well as the reconfigurable coexistence of superconducting and normal states, was demonstrated. As an alternative material, 3D niobium-carbon (Nb-C) structures, such as interconnected wires[52] and scaffold bridges[58] have also been grown via $Ga^+$ FIBID, and exhibit $T_c$ values around 6 K, comparable to their planar counterparts.

When considering the difference between 2D and 3D depositions, both W-C[54,96,97] and Nb-C[52] planar structures fabricated via FEBID and FIBID exhibit lower $T_c$ values, typically below 5 K. Nevertheless, sub-100 nm planar W-C structures have revealed remarkable phenomena, including vortex confinement [98], non-local vortex transfer[99,100] and superconductivity controlled by electric fields[101], showcasing their promise in advanced superconducting devices.

As with the magnetic materials, there are a number of superconducting materials that have been exclusively grown in 2D, including molybdenum-carbon (Mo-C) nanostructures with $T_c$ < 4 K[102,103], and lead-carbon (Pb-C) deposits exhibiting $T_c$ = 7.2 K[104], as detailed in Table 1. Despite these advances, the number of superconducting materials grown in 3D at sub-100 nm scales remains limited. This limitation underscores the need for interdisciplinary collaboration, particularly with chemists, to design tailored precursor molecules capable of achieving the required desired purity levels, either directly or through appropriate post-treatment strategies, while also remaining compatible with FEBID/FIBID workflows and environments, which is not straightforward.

## Future developments of the technique

*Spatial resolution*

Given that the primary electron beam spot and the individual molecules are ∼ 1 nm in size, planar dots as small as this value have been demonstrated[105], setting a benchmark for this technique. However, when applied to 3D structures, achievable resolutions are typically in the range of a few tens of nanometres, with feature sizes ≲ 10 nm possible at structure edges or boundaries[19]. Improving printing resolutions of complex 3D objects further is challenging, since the focused beam continuously interacts with the growing structure throughout the entire printing process. While the 3D printing-inspired layer-by-layer approach[12] (see **Box 1**) allows for the creation of arbitrary 3D geometries, changes in the interaction volume as the structure builds up require specific compensation. This leads to feature enlargement and unintentional increases in the dimensions of the intended geometry. Addressing this effect will require new procedures compensating for these cumulative interactions, at the cost of more complex algorithms and computational running times. While these adjustments could improve accuracy and fidelity when growing complex nanostructures, the challenge lies in developing computationally feasible solutions that do not compromise the current workflow's efficiency and simplicity – something that could benefit from the use of machine learning techniques.

*Broader materials and functionalities*

A distinguishing feature of FEBID/FIBID compared to other 3D nanoprinting techniques is their unique ability to directly pattern functional nanomaterials using specific precursor molecules[19,19,21,27,106] (in this article focused on ferromagnets and superconductors). However, a main challenge that limits their broader application is the limited variety of available materials. Notable progress has been made with the development of heteronuclear precursors[107], with the deposition of Co-Fe and Co-Si alloys as initial demonstrations[108–110]. As highlighted in Section 2, there are a number of precursors available for which 3D growth has not yet been explored. Going forward, close collaboration with synthetic chemists would provide a route to design new precursors[107], allowing the community to target different scientific questions and phenomena. One could consider materials with non-trivial intrinsic symmetries, as well as superconductors with longer coherence lengths or higher critical temperatures.

There are several alternative approaches to achieving a broader range of materials: the use of a multi-gas or a multiple single-gas GIS to mix several precursors and reactive gases [84,111,112], as well as the development of post-processing methods like global[113,114] and local[58,115,116] annealing, and electron/ion irradiation[114,117].

For the realisation of layered heterostructures, key for many applications, deposition with multiple precursors has been demonstrated [70,118]. With typical growth rates of 10 nm/s to 100 nm/s, high vacuum conditions in the microscopy chamber, and normally a lack of pressure feedback at the sample area, the formation of optimal interfaces is challenging. While this has been demonstrated in a proof-of-principle manner for a magnetic tunnel junction[70], extending such interface control more broadly remains non-trivial. A well-established alternative to directly patterning magnetic or superconducting materials using FEBID and FIBID is the implementation of hybrid multi-step fabrication approaches. Here, the printing of a 3D scaffold is first performed, followed by the deposition of a thin film on top, either via physical or chemical method, to cover the nanostructure[38,57–59,119]. The 3D geometry is thus established by the printed scaffold, while the material's functionality is provided by the thin film deposited on top. Physical deposition methods can excel in creating multilayered materials, although they often face challenges such as shadowing effects that can inhibit uniform coverage. In contrast, chemical deposition methods tend to be largely conformal, providing more consistent coverage over complex geometries. We note that issues such as high temperatures and material compatibility can introduce additional complications in the fabrication process, which typically need to be addressed for each application. With dedicated efforts, the realisation of a robust approach to multi-step fabrication processes will open the door to the realisation of functional 3D heterostructures.

*Speed and scalability*

While FEBID and FIBID currently offer the best spatial resolution available for 3D printing, they are inherently slow techniques due to their sequential nature. As a result, their primary real-world applications to date have been lithography mask repair[120,121] and the fabrication of highly customized scanning probe microscopy tips[122,123]. In both cases, a low throughput is viable; however, many other possible applications will ultimately require substantially higher fabrication rates. This scalability challenge could be addressed through several potential routes. First, multi-beam tools offered from a few manufacturers for large-scale electron beam lithography[124–126] could be adapted for FEBID/FIBID. A second, more accessible method is the use of cryogenic temperatures (Cryo-FEBID/FIBID)[127], where a multi-layer adsorbate leads to a manifold increase in growth rates[128,129]. This second route remains relatively unexplored, with current efforts aimed on preserving the functional properties of the resulting materials. New developments in FEBID modelling will help to progress in this realm[76]. Finally, recent advances in patterning planar metallic deposits through ion irradiation of organometallic thin films show potential for direct writing of planar metallic interconnectors[130,131]. However, extending this approach to 3D would require a way to confine this process to a local 3D voxel inside a thicker medium, analogous to the optical case of two-photon lithography[132,133].

*Robustness of technique*

An additional challenge to be addressed after the recent development of nanoscale 3D printing capabilities is the reproducibility of these processes, especially when working with new precursor gases and particularly complex geometries. Indeed, it is highly important that the growth and realisation of nanostructures be transferrable and reproducible between instruments and groups in the community. Dedicated studies aimed at reproducing outcomes under varying conditions will be essential to establish reliable protocols and methodologies. A key aspect for all these applications lies in the composition of the resulting materials, with the remaining carbon content playing a significant role in their properties. For instance, the elastic constants may vary by several orders of magnitude depending on the carbon/metal ratio within the composite material[134], radically changing the mechanical characteristics of a nano-device[19]. This variability highlights the tunability of FEBID/FIBID materials, with the possibility to target different compositions for different applications. This potential to exploit 3D geometries for varying applications – for example, targeting coupling between mechanical and magnetic or superconducting properties, or the integration of other FEBID/FIBID-compatible materials e.g., plasmonic materials[19,111,135–138] – in a single device is a young yet promising area of research[139] with numerous opportunities.

## 3. Future perspectives

The development of focused electron and ion beam induced deposition for 3D nanoprinting has brought in a new era for magnetic and superconducting nanostructures, not only allowing for the exploration of geometric effects and resulting emergent phenomena, but enabling the fabrication of first prototype devices for potential applications. Now that these techniques are established, the community can look forward to a maturing field, and with it, the resolution of outstanding challenges. For example, with interdisciplinary collaborations, the identification and development of appropriate precursors will extend 3D nanoprinting to a wider variety of materials with more exotic intrinsic properties and symmetries, allowing new physical phenomena to be explored. The move towards three-dimensional heterostructures (and the optimisation of the interfaces within), which requires the development of multi-deposition processes, will open up possibilities for three-dimensional spintronic, or superconducting spintronic devices, ushering in a new generation of technological applications. Finally, to move from prototypes to industrial devices, an upscaling of three-dimensional nanoprinting is crucial – something that would require either enhancing the growth rate of FEBID/FIBID, for example by exploiting cryogenic growth or multi-beam setups – or by developing alternative industry-compatible method for specific geometries identified by FEBID/FIBID prototyping.

In the coming years, we foresee a multitude of possibilities for further development. For example, combining the functional and mechanical properties of three-dimensional nanostructures – by integrating FEBID/FIBID grown magnetic and superconducting nanostructures into MEMS and NEMS devices – could open up possibilities for three-dimensional magnetomechanical nanostructures[139]. Such combinations could offer strain control of emergent properties, with many possibilities for applications ranging from magnetomechanical sensing devices, to nanorobotics.

From a technological point of view, one of the most motivating aspects of three-dimensional architectures is the prospect for increased interconnectivity. Realising highly interconnected three-dimensional devices with FEBID/FIBID currently faces the challenges of parasitic deposition and inhomogeneous growth with height. Further developments to control complex depositions –by harnessing different growth regimes or postprocessing of the grown architectures[140] – could benefit from the incorporation of advanced computational methods[141] such as artificial intelligence into the modelling of the growth and processing.

Finally, we foresee the exploration of new physical phenomena via the patterning of a wider range of materials as one of the most exciting next steps in the field of FEBID/FIBID. For example, achieving precise crystalline materials such as chiral magnets requires not only the development of new precursors, but fine control over the composition and microstructure of the deposits. Once achieved, we expect the exploration of competing extrinsic (geometrical) and intrinsic (material) effects to provide routes to control – and indeed entirely new – fundamental phenomena.

**BOX 1: Details of FEBID/ FIBID:**

Focused Electron Beam Induced Deposition (FEBID) and Focused Ion Beam Induced Deposition (FIBID) are typically conducted in a scanning electron microscope (SEM) / focused ion beam (FIB) microscope, equipped with a gas injector system (GIS) and a pattern generator[11,21]. Here, a high-energy electron or ion beam, usually in the kilovolt range, strikes the substrate, enabling nanoscale deposition on the substrate, in the region where the beam is scanned. The GIS introduces the precursor gas needed for deposition. It consists of a heated reservoir that produces gas from a solid or liquid precursor. This gas flows through the GIS to a nozzle located close to the substrate, where it adsorbs onto the surface. As the beam scans over this surface, it decomposes the adsorbed gas molecules, creating a localised deposit. FEBID/FIBID resembles chemical vapor deposition (CVD), but the reaction is primarily driven by the focused beam rather than heat (which may however play a key role in some cases[12,144,145]). The pattern generator, operated through a digital-to-analog converter (DAC), enables accurate control of beam coordinates and dwell times. These are then implemented with high precision through the use of an electrostatic blanker to direct the beam, allowing for adjustments in dwell and refresh times down to a few tens of nanoseconds. FEBID/FIBID is thus a direct-write additive nano-manufacturing technique without the need of templates or resists. An additional bonus is that nanofabrication is in principle possible on even the most challenging substrates[135], provided excess charge is dissipated to prevent charging effects.

Spatial resolution in FEBID/FIBID is typically in the range of a few tens of nanometres, and comes determined mainly by the interaction volume between the beam and the substrate[146]. When the beam strikes the surface, it loses some of its energy, generating a wide range of particles, such as lower-energy electrons and X-rays. Among these, low-energy secondary electrons are particularly effective in decomposing precursor gases, as they have a high interaction cross-section with adsorbed molecules[21] with several pathways possible for decomposition[147].

FEBID and FIBID are inherently very complex techniques, with numerous competing effects, diverse timescales, and intricate interactions of electrons with matter in both solid and gas phases that may change over the growth of a nanostructure[148]. Direct control over the fundamental physical parameters is not possible. Instead, it is possible to influence multiple growth parameters by changing experimental parameters such as current, voltage, dwell and refresh times, base and growth pressures, patterning strategies, and GIS geometry. All these collectively influence the 3D structure, material composition, and properties of the printed nanomaterials[19]. To deal with this complexity and enable 3D printing, a range of computational tools are being developed. These include Monte Carlo simulators for modelling both GIS gas flow[149] and beam-matter interactions[150–152], FEBID/FIBID continuum models that account for growth kinetic processes like adsorption, desorption, diffusion and dissociation[76,153,154], and molecular dynamics models combined with density functional theory (DFT)[155,156] for deeper insights at the molecular level. Among these computational tools, two main approaches have emerged for 3D printing with FEBID. The first approach that was demonstrated[13,48] (Fig. B1 (a-f)) constructs “hollow” or “meshed” structures by fabricating networks of interconnected nanowires, with the possibility for these to be extended to connected planes[49]. This approach relies on experimentally relating the growth angle of a nanowire with the speed of the beam during patterning (Fig. B1 (b-d)). By using comprehensive simulations[157] involving the FEBID continuum model and gas dynamics (Fig. B1 (a)), and finding the parameters that reproduce the angle-vs-speed relationship obtained in experiments (Fig. B1 (c)), subsequent 3D growth of complex meshed nanostructures becomes possible ((Fig. B1 (d)). Given that most of these parameters are difficult to obtain experimentally, many fitting parameters are required in simulations[144], which may result in degeneracy in the parameter space. This approach excels in the growth of complex 3D nanowire-based geometries (Fig. B1 (e,f)) but is not currently suited for other types of shapes.

A second, recently demonstrated, approach[12] (Fig. B1 (g-i) is more similar to the one followed by a standard 3D printing extruder. This approach enables for the first time the fabrication of arbitrary nano-geometries, including not only nanowire-based, but also surfaces and volumes (Fig. B1 (h,i)). This framework leverages the FEBID continuum model under the approximations of stationary conditions and/or long dwell times, together with limited diffusion. Under these conditions, typically present in regimes where high-fidelity 3D printing is achieved, the number of effective parameters can be reduced to only a few (typically two or three), namely growth rate, interaction volume, and a heating correction factor. A two-step calibration experiment first determines experimentally these parameters. Afterwards, the geometry is sliced in multiple layers, and the aforementioned model efficiently computes the dwell times for each layer. A few additional effects *e.g.*, anisotropic gas flux and beam defocusing may also be incorporated into the model after additional calibration experiments. Fig. B1 (g) shows an overview of the algorithm. Thanks to the recent developments for 3D nanoprinting, FEBID/FIBID 3D printing offers the ability to directly write complex 3D nanostructures based on functional materials (e.g. magnetic and superconducting), with a high degree of precision and versatility. Recent efforts to popularise FEBID/FIBID for nanoscale 3D printing include free access to the 3D printing codes, online tutorials[158], the development of a unified platform to integrate existing software and connect research groups[159], demonstrated compatibility with various precursors, and code portability for use with microscopes from major manufacturers[158].

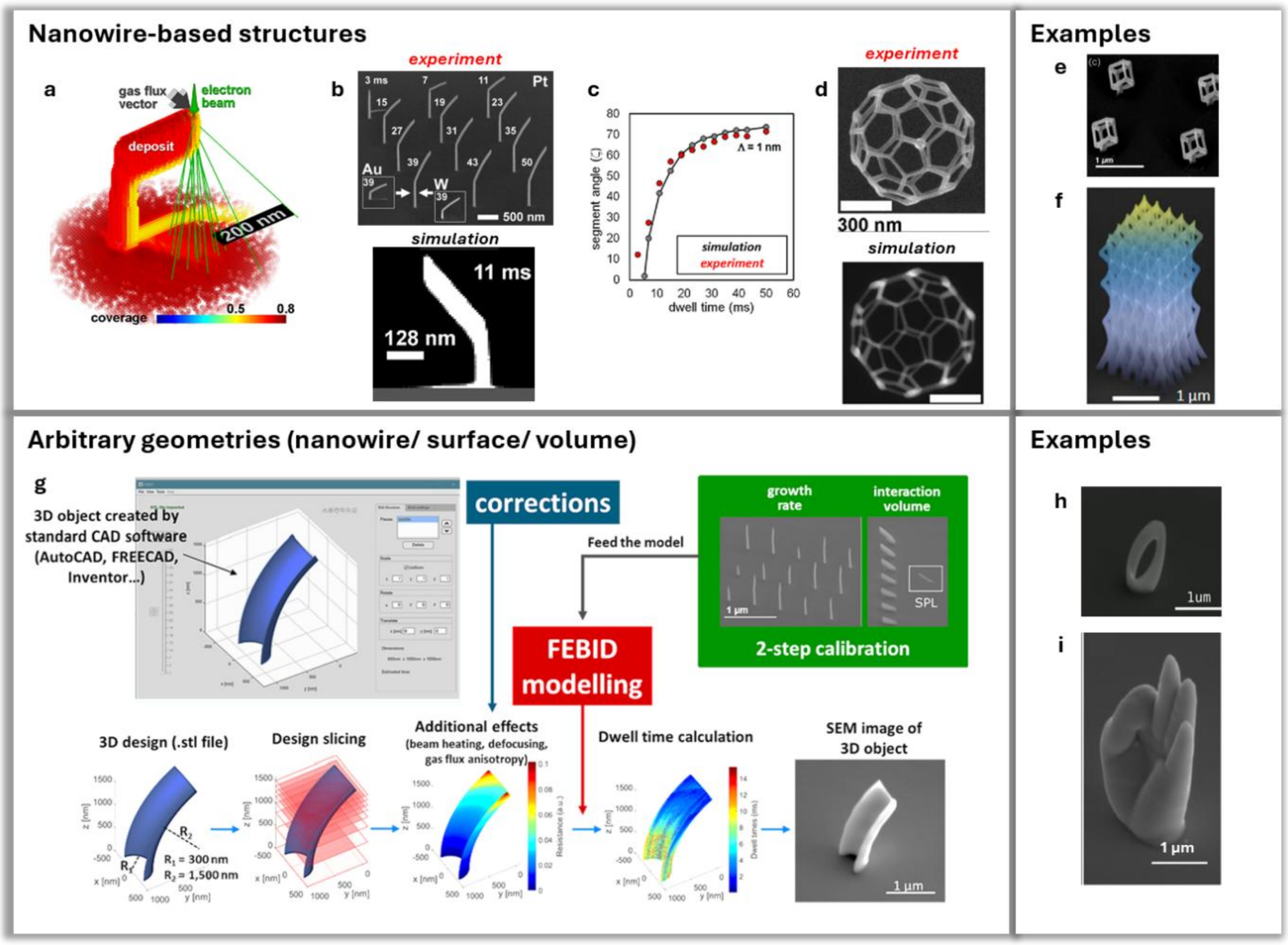

*Figure B1.* Two current main methods for 3D nanoprinting by FEBID (readily extendable to FIBID). Top (a-f): First approach, based on the growth of meshed, hollow structures, formed by interconnected nanowires. (a) Snapshot of a combined FEBID continuum model- electron-matter interaction Monte Carlo simulator to model the growth of nanowires fabricated at an angle with respect to the substrate, after the growth of a vertical pillar. (b) Experimental study (left) of such nanowires with different dwell time per point used at each spot, for a 1 nm pitch in between points. The simulation (right) shows how the parameters used during simulations leads to a good agreement in the case of a 11 ms dwell time. (c) Good agreement between experiments and simulations demonstrated for a wide range of dwell times, enabling the fabrication of complex meshed 3D nanostructures. (d) Experimental realization (top) and corresponding simulations (bottom) of a 3D truncated icosahedron, as an example of a complex meshed nanostructure realised using this approach. (e-f) Examples of meshed nanowire structures (cubes and multi-layered hollow object) using this methodology. (g-i): Second approach, based on a layer-by-layer 3D printing of arbitrary shapes (nanowire-based, surfaces and volumes). (g) Schematics of the algorithm used by the open-access f3ast for 3D nanoprinting by FEBID. Starting from a 3D CAD object in a .stl format, the software: imports the file; slices the geometry; takes into account possible effects like heating, defocusing and gas flux anisotropy; calculates the necessary dwell time per point; generates the streamfile read by the DAC of the pattern generator of the microscope to build the structure; the final SEM image on the right is the result of this algorithm, in this case a double-curved nano-surface. The FEBID model used to calculate the dwell times is based on the continuum model and requires at most three inputs from a 2-step calibration process. The first one consists of growing a series of vertical nanowires, to determine the growth rate; the second grows nanowires at an angle with respect to the substrate, where the angle formed by the substrate is compare with a single pixel line (SPL) nanowire, to determine the effective interaction volume of the beam between points in a given layer. (h-i) Examples of surface (Möbius strip) and volumetric (nano-hand) patterned using this approach. Images in (a-d) reproduced from [157] with permission. Image in (e) reproduced from [48] with permission. Image in (f) reproduced from [135] with permission. Images in (g-i) reproduced from[12] with permission.

**BOX 2: Measurement of 3D SC and MAG nanostructures**

One of the key aspects of nanopatterning functional materials is that we can gain control over their emergent properties. However, many established characterisation techniques are designed for planar structures. This creates an experimental challenge: how to measure and interpret these emergent properties in 3D – both to understand the behaviour, as well as providing routes to reading information from future devices.

*What are we trying to measure?*

For magnetic nanostructures, we probe *e.g.,* the domain configuration, the presence of magnetic defects, as well as spintronic effects. For superconducting nanostructures, we probe *e.g.,* the SC or N states in the nanostructure, the presence of defects such as superconducting vortices, and quantum mechanical effects such as the Aharonov-Bohm or Josephson effect.

Measurement of the state and configuration can be achieved through direct measurement of the magnetisation (with optical or X-ray techniques), the measurement of local stray fields (electron and scanning probe techniques) or through transport – where one probes spintronic effects, or changes in resistance due to the superconducting state (for example, due to phase slips). In the following, we will briefly discuss these different measurement routes, with a focus on the how they may be adapted to three-dimensional nanostructures grown with FEBID.

***Nanoimaging:* Magnetisation (3D-FM)**

In a magnetic material, the magnetisation can be directly probed with photon-based techniques. First, with X-ray magnetic circular dichroism (XMCD), element-specific mapping of the component of the magnetisation parallel to the beam can be achieved with nanoscale spatial resolution, with the potential for MHz and GHz stroboscopic imaging. When combined with tomographic measurement geometries, this approach allows X-ray magnetic tomography[160–163], first developed with bulk[161] and more recently applied to three-dimensional magnetic nanostructures[78,164]. In this way, the three-dimensional vector field, and the effect of 3D nanogeometries, can be directly determined. While the superconducting order parameter cannot be directly probed with XMCD, imaging the magnetic configuration of soft magnetic materials deposited on superconductors can allow insight into their nanoscale configuration[165–167].

Access to the magnetisation can also be obtained with optical techniques in the visible range via the magneto-optical Kerr effect (MOKE). Although limited in spatial resolution, dark-field MOKE geometries can be used to measure the switching of individual components within 3D nanostructures, such as nanowires oriented at different angles with respect to the substrate[38]. The use of optical techniques also allows for the probing of dynamic processes in three-dimensional geometries, using ultra-fast pump probe MOKE[168], as well as the probing of magnons with Brillouin light scattering[34].

***Nanoimaging:* Magnetic induction (3D-FM, 3D-SC)**

Both superconducting and ferromagnetic materials can be investigated with stray field mapping. In a ferromagnet, the magnetisation contributes to B within the sample, and inhomogeneities such as magnetic textures act as a source of B outside the sample. In superconductors, the superconducting state manipulates the local magnetic field through the Meissner effect, allowing for contrast between superconducting and normal regions of the sample, and for enhanced stray field in the vicinity of superconducting vortices.

Probing the stray field can be achieved in transmission using transmission electron microscopy (TEM). With Lorentz microscopy and electron holography, it is possible to recover a high spatial resolution map of the components of the magnetic induction in the plane perpendicular to the magnetic field. Such imaging has been applied to image *e.g.,* the stray field and magnetic configuration in the vicinity of FEBID Co nanohelices[32]. Three-dimensional vector maps of the magnetic induction have been obtained using tomographic techniques[169,170], including within a FEBID Co nanowire[171]. While TEM imaging of 3D superconducting nanostructures has not yet been realised, imaging of vortices has previously been achieved in flat samples[172,173], setting a precedent for the extension to three-dimensional imaging of superconducting configurations.

Maps of the magnetic induction or its gradients can also be obtained in the vicinity of the sample surfaces with scanning probe techniques such as magnetic force microscopy (MFM) and scanning nitrogen vacancy magnetometry. For MFM, high spatial resolution of the local stray field can be obtained in the vicinity of the sample. However, the magnetic tip may perturb the sample due its magnetic moment. The more recently established technique scanning NV offers ultra-sensitive measurements of weaker magnetic fields in a non-invasive manner, with the possibility to extend to time resolved GHz measurements[174].

***Transport:* Electrical transport (3D-FM, 3D-SC)**

In addition to microscopy techniques, insight into the properties of 3D ferromagnetic and superconducting nanostructures can be obtained via transport experiments. Beyond sample characterisation, transport measurements are particularly relevant for future devices, which typically are designed with electrical read/write in mind, and can also reveal emergent transport effects arising from geometry and topology.

For ferromagnets, transport measurements harness spintronic effects to provide signals that can be proportional to the orientation of the magnetisation, the presence of defects, or the population of thermal magnons in a sample[77]. For superconductors, transport measurements provide a direct electrical signature of the superconducting state: ideally, zero resistance reflects a fully phase-coherent condensate, whereas a finite resistance can arise from dissipation mechanisms such as vortex motion[17] or phase slips[18], as well as from pair-breaking processes that generate quasiparticles and weaken superconductivity.

One of the main challenges in transport measurements is integrating three-dimensional nanostructures into electrical circuits. With FEBID and FIBID, direct writing of structures onto two-dimensional electrical contacts can represent a simply route to measuring transport. However, the parasitic deposition that is typical of FEBID and FIBID structures can introduce unwanted conduction paths, posing a challenge that can be mitigated by modelling[77] or by prestructuring the substrate[17,57]. Alternatively, micromanipulation of the samples onto electrical contacts can provide a cleaner measurement[18], albeit limited when it comes to small structures or complex geometries. Finally, beyond conventional contacting, growing nanostructures directly on top of lithographically patterned high-sensitivity Hall probe devices enables measurement of their magnetic switching behaviour[175].

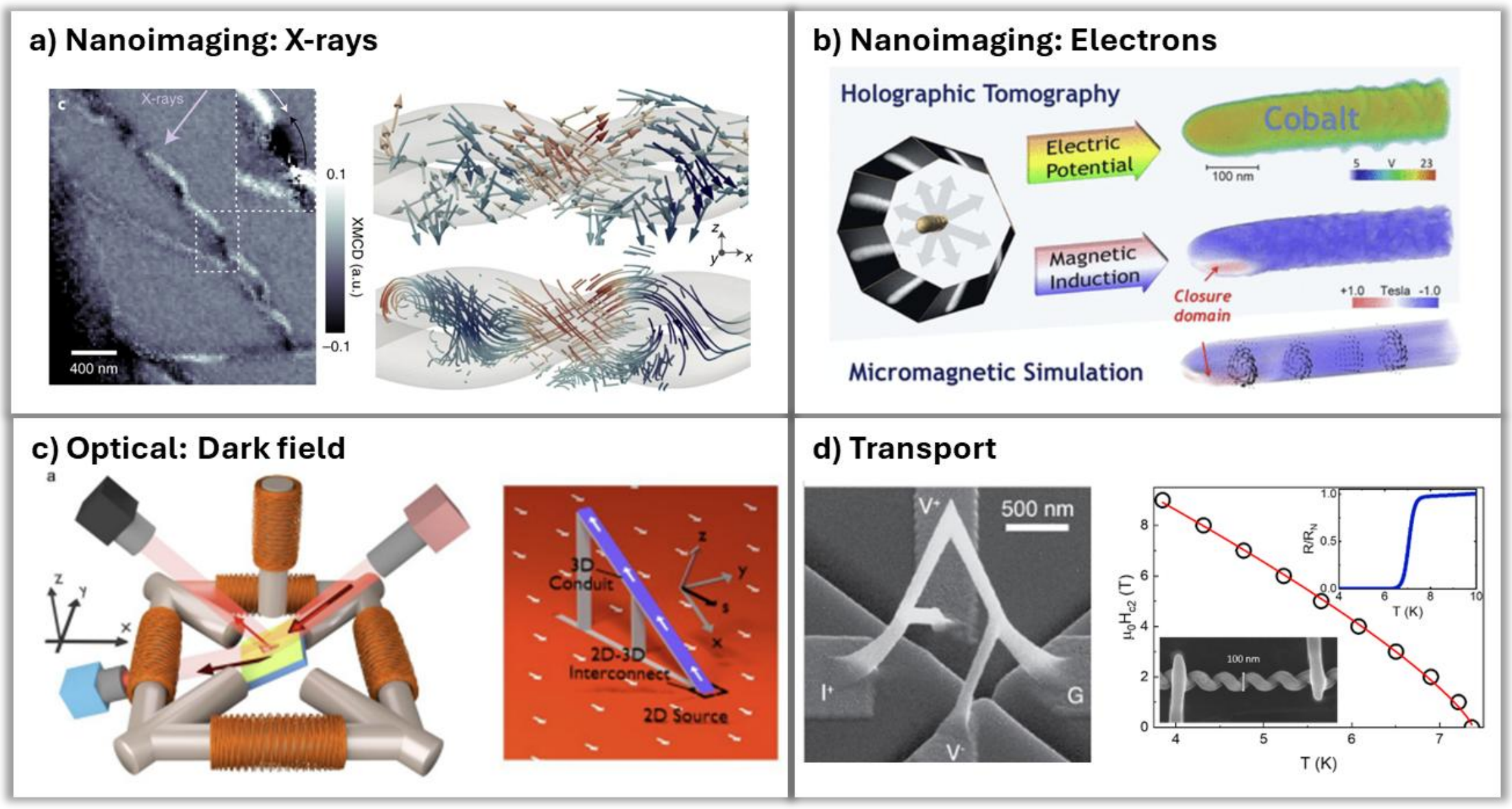

Figure B2. Selection of characterisation techniques used to experimentally investigate 3D FEBID/FIBID nanostructures. a) The three-dimensional magnetic configuration of magnetic nanostructures can be investigated using X-ray magnetic circular dichroism in combination with high spatial resolution nanoimaging. Two-dimensional projections that provide the integration of the magnetisation along the beam provide insight into the configuration, while acquiring many such projections for different orientations of the sample allows for the determination of the three-dimensional magnetisation vector field with X-ray magnetic tomography. b) Imaging of the magnetic induction with transmission electron microscopy using, for example, Lorentz microscopy or electron holography, allows for the investigation of both magnetic and superconducting nanostructures. As with X-rays methods, acquiring 2D projections in a tomographic geometry allows for the mapping of the three-dimensional magnetic vector field. c) Magneto-optical Kerr effect magnetometry allows one to probe the magnetic switching of a sample. Using an adapted "Dark-field" geometry, the reflected light from an angled 3D nanostructure can be spatially separated from that of the substrate, allowing for the direct measurement of the switching processes of individual 3D nanostructures. d) By integrating magnetic and superconducting nanostructures onto electrical contacts, one can measure transport. For magnetic nanostructures, this provides insight into the magnetic configuration and switching behaviour, while for superconducting nanostructures, the transport measurements provide insight into the superconducting transition, the presence of phase slips and vortices, and emergent quantum effects. a) reproduced from [78]. b) reproduced from [75]. c) reproduced from [38]. d) reproduced from [18,77].

**TABLE 1: Materials and associated properties available with FEBID and FIBID**

| | | |
|---|---|---|
| **Magnetic:** | 3D | 2D |
| **Superconducting:** | 3D | 2D |

| Metal | Precursor | Resolution (nm) | Composition (atomic %) | Crystallinity | Magnetism / Superconductivity | Comnents | Ref. |
|---|---|---|---|---|---|---|---|
| **Co** | $Co_2(CO)_8$ | ~50 - 70 | Co:C:O 59-97 : 2-41 : 1-8 | 3–6 nm crystal size<br>Co *hcp* | $\mu \approx 10^{-11}$ emu, $\lambda_{ex} = 4 \pm 1$ nm, $M_s$ = 1190 emucm$^{-3}$ | FEBID | 12,28,56,57,72–79,176 |
| **Fe** | $Fe(CO)_5$<br>$Fe_2(CO)_9$ | 30 – 70 | Fe:93 O:1 C:6; Fe:60 | α-Fe, $Fe_3C$, $Fe_5C_2$, and $Fe_7C$ | $M_s \approx 100$ kA m$^{-1}$ | FEBID, heat post-treatment | 73,80–82 |
| **Co-Fe** | $HFeCo_3(CO)_{12}$ | 80 - 110 | FeCo:C:O 80:10:10 | bcc $Co_3Fe$ phase | $T$= 25 K, $\mu_0 H_{switching}$=14 mT | FEBID | 50,58,83 |
| **Fe-Pt** | $Fe(CO)_5$ + $(CH_3)_3CH_3C_5H_4Pt$ | 26 | Fe:Pt 50:50 | (*fct*) FePt phase ($L1_0$, P4 / mmm) | Br= 1.53 T | FEBID, heat post-treatment | 84 |
| **Co** | $Co_2(CO)_8$<br>$Co(CO)_3NO$ | 10 – 1500 | Co:C:O 90:8:2<br>Co:C:Ga:N:O 54:7:9:13:17<br>Co:C:N:O 50:10:14:26 | 5- 20 nm crystal size<br>Co *hcp*, CoO *fcc* | $H_c \approx$ 50–200 Oe; $M_s \approx$ 400–1200 emu/cm$^3$; AMR = -0.25 - -0.9 %; $\rho_H$= 2.7 µΩcm; $\phi_{min} = 2 \times 10^{-6}\ \phi_0$; $S_I$ = 0.13 Ω/T | FEBID; $Ga^+$ FIBID; $He^+$ FIBID; Ga-cryo-FIBID<br>Heat post-treatment | 177–181 |
| **Fe** | $Fe(CO)_5$<br>$Fe_2(CO)_9$ | 60 - 1000 | Fe:C:O 76:12:12 | amorphous matrix<br>~1–2 nm crystal size α-Fe | MR= -0.59 %; $M_S$= 1371 kA m−1; ρAHE= 12 µΩ cm | FEBID, Heat post-treatment | 112,182–184 |
| **Co-Pt** | $Co_2(CO)_8$ + $(CH_3)_3CH_3C_5H_4Pt$ | 1000 | Co-Pt:C:O 25:60:15<br>*Co-Pt:C:O 38:39:23 | * fct phase ($L1_0$) nanocrystals | Superparamagnetic; * Ferromagnetic<br>$H_c$ = 292 Oe, $M_r/M_s$ = 0.11 | FEBID<br>*electron irradiation treatment: | 85 |
| **Co-Si** | *$Co_2(CO)_8$ + $Si_5H_{12}$<br>$H_3SiCo(CO)_4$<br>$H_2Si(Co(CO)_4)_2$ | 1000 | *Si:C:O:Co 28-47:23:15:24<br>Co-Si:C:O 55:30:15 | CoSi; $CoSi_2$ | $M_s$= 1.5 T; $MR_{CoSi}$ ~ 0.7 %; $MR_{Co2Si}$ ~ 1 % | FEBID, *tuning by precursor flux | 86,109 |
| **Co-Fe** | $HFeCo_3(CO)_{12}$ | 50 | FeCo:C:O 80:10:10 | bcc Co-Fe phase + spinel oxide $FeCo_2O_4$, 5 nm crystal size | $M_s$= 1.2 – 1.6 T; $\sigma_{xx}$= $2.3 \cdot 10^4\ \Omega^{-1}$ cm$^{-1}$; $\sigma_{xy}$= $1.9 \cdot 10^2\ \Omega^{-1}$ cm$^{-1}$; MR < 1%; $\mu \approx 5 \cdot 10^5$ cm$^2$V$^{-1}$ S$^{-1}$; n≈ $2.5 \cdot 10^{11}$ cm$^{-2}$ | FEBID | 108 |
| **Co-F** | $Co_2(CO)_8$ + $XeF_2$ | 1000 | $CoF_3$ phase | polycrystalline grains of nanometric size | Not reported | FEBID, FEB irradiation + $XeF_2$ | 88 |
| **Ni-O** | $(MeCp)_2Ni$ | 1000 | Ni:O:C 50:40:10 | polycrystalline 5 nm size, NiO fcc | Not reported | FEBID, FEB irradiation + $O_2$ | 89 |
| **Fe-Si** | $Fe(CO)_5$, $Si_5H_{12}$ | 2000 | Fe:Si:C:O 31:37:14:19 | *B*20 phase of *FeSi*<br>*D*0$_3$ crystal structure of $Fe_3Si$ | Not reported | FEBID, FEB irradiation + heating, thermal post-treatment | 90 |
| **$Co_2FeSi$** | $HFeCo_3(CO)_{12}$ + $Fe(CO)_5$, $Si_5H_{12}$ | 2000 | Fe:Co:Si:C:O 17:33:17:15:17 | Phase ($L2_1$) $Co_2FeSi$ | Not reported | FEBID, FEB irradiation + heating | 91 |
| **W** | $W(CO)_6$ | ~32 – 100 | W:C:O :20-8:8-7 | 3 - 30 nm crystal size, fcc WC | $T_C$ = 4.7 – 7.1 K; $B_{c2}$ (0 K)= 9.7 - 15 T<br>ξ= 4.86 – 5.75 nm; λ= 830 – 1200 nm | $He^+$ FIBID, $Ga^+$ FIBID, FEBID | 17,18,29,92–95 |
| **Nb** | $Nb(Nme_2)_3$(N-t-Bu) | 170 | Nb:C:O:N 20-25:35:20–25 | 15 - 20 nm-sized nanocrystals, fcc NbC | $T_C$ = 6.1 - 8.1 K; $B_{c2}$ (4 K)= 8 – 31 T; ξ= 4 - 5.85 nm; λ= 471 - 775 nm | $Ga^+$ FIBID, CVD | 52,58 |
| **W** | $W(CO)_6$ | ~20 – 50 | W:C:O:Ga 50:30:10:10<br>W:C:O 50:40:10 | short-range nano-crystalline order | $T_C$ = 2.0 – 5.7 K; $B_{c2}$ (0 K)= 3.7 – 9.5 T<br>ξ= 6.25 nm; λ= 616 – 850 nm | $Ga^+$ FIBID, $He^+$ FIBID, FEBID | 54,96–98,100,185–190 |
| **Nb** | $Nb(Nme_2)_3$(N-t-Bu) | 500 | Nb:C:Ga:N 29:43:15:13 | 15 - 20 nm-sized nanocrystals, fcc NbC | $T_C$ = 5 K; $B_{c2}$ (0 K)= 9 T; ξ= 5.94 nm;<br>λ= 1101 nm | $Ga^+$ FIBID | 52 |
| **Mo** | $Mo(CO)_6$ | ~50 | Mo:C:Ga:O 41:26:26:7<br>Mo:C 70:30 | Amorphous | $T_C$ = 2.7 – 7.2 K; $B_{c2}$ (0 K)= 2.5 – 4.5 T;<br>ξ= 7.6 – 11.2 nm | $Ga^+$ FIBID, FEBID + with water vapor | 102,103 |
| **Pb** | $(CH_3CH_2)_4Pb$ | 1000 | Pb:C:O 46:31:23 | Not reported | $T_C$ = 6.6 - 7.2 K; $B_{c2}$ (0 K)= 9.9 – 0.20 T;<br>ξ= 5.75 - 40.23 nm | FEBID, Two superconducting phases | 104 |